\documentclass[12pt]{article}
\usepackage{graphicx}
\usepackage{float}
\usepackage{amssymb,amsmath,amsfonts,palatino,amsthm}
\usepackage{amssymb}
\usepackage{epstopdf}
\usepackage{slashed} 
\newcommand{\f}{\begin{equation}}
\newcommand{\ff}{\end{equation}}

\begin{document}

\title{Holographic relations in loop quantum gravity \\}
\author{Lee Smolin\thanks{lsmolin@perimeterinstitute.ca} 
\\
\\
Perimeter Institute for Theoretical Physics,\\
31 Caroline Street North, Waterloo, Ontario N2J 2Y5, Canada}
\date{\today}
\maketitle

\begin{abstract}
\end{abstract}

It is  shown that a relation between entropy and minimal area holds in loop quantum gravity, reminiscent of the Ryu-Takayanagi relation.  

\tableofcontents

\newpage

\section{Introduction}

The Ryu-Takayanagi relation\cite{RT} is one of the deeper expressions of holography in $CFT$ and condensed matter theory.  It relates the entanglement entropy across a boundary $\gamma$ in a conformal field theory to the minimal area of a surface,
$\sigma$ that has boundary $\gamma$ but suspends into a bulk of one higher dimension.  It has stimulated very fruitful developments concerning the
relationship between space-time geometry and entanglement\cite{AdS-grav-entanglement}.  It suggests a new way to understand entanglement entropy\cite{Rafael-entbh} as well as the older results relating 
thermodynamics, spacetime and the quantum\cite{jb}-\cite{cs}.

The Ryu-Takayanagi relation is expressed in terms of a diffeomorphism invariant observable
in the bulk-an extremal area, which raises the natural question of whether it can be formulated, or has an analogue, in diffeomorphism invariant approaches to quantum gravity.  In this paper, I show one such partial analogue, in the context of loop quantum gravity, and conjecture another, closer analogue.

Loop quantum gravity provides a natural setting for a bulk/boundary correspondence, which was pointed out in \cite{linking} and has been developed in many papers since\cite{Kirill-bh,isolated,aldoetal,others}.  

This arises from the fact that  general relativity has a close connection to a topological field theory\cite{Plebanski,CDJ}.  Specifically, as we review in the next section, general relativity arises from the topological $BF$ theory by the imposition of certain constraints, called the {\it simplicity constraints\cite{BC,FK,EPRL,reviewSF}.}  Further, there is a natural 
{\it ladder of dimensions}\cite{Louis1} which relates $BF$ theory on a four dimensional  manifold $\cal M$ to Chern-Simons theory on its boundary,
${\cal B}= \partial {\cal M}$.   The combination of these circumstances gives a boundary theory related to Chern-Simons theory for classical and quantum general relativity\cite{linking}.  

This bulk/boundary relation has been central to the description of black hole and cosmological horizons in loop quantum gravity, where the space of states on the horizon is related to Chern-Simons theory\cite{linking,Kirill-bh,isolated}.  However the full importance and meaning of this relation remains to be elucidated.  The relation is particularly simple in the cases of non-vanishing cosmological constant, 
$\Lambda$, and this suggests these structures may enable a background independent realization of an $AdS/CFT$ correspondence, or even lead  to a general notion of holography.  This was suggested in \cite{linking} but the exact sense in which this provides a realization of holography\cite{tHooft-holo,Lenny-holo}, or might be related to the $AdS/CFT$ correspondence\cite{Malda1}, remains obscure.

Happily,  a new piece of the puzzle has recently come to light.  This is the realization that the simplicity constraints express the first law of thermodynamics\cite{carloetal,FGP,Eugenio1,thermost}, and are hence closely related to black hole thermodynamics\cite{FGP,Eugenio1}. Through this they are related to a key relation that holds on the corners of  causal diamonds, called by some the {\it first law of spacetime dynamics}\cite{CT,MP,BW,ls-bh,FGP,Eugenio1}. This has yielded insights into black hole  entropy\cite{FGP,Eugenio1},as well as enabling a realization of Jacobson's argument that the Einstein equations reflect the thermodynamics of an underlying pre-geometric physics\cite{Ted95,Paddy,Ted2015}, within loop quantum gravity\cite{ls-bh}.  Here we show that this additional insight, when combined with the older results, enables us to derive an analogue of the Ryu-Takayanagi relation within the framework of loop quantum gravity.

In the next section we review the basics of loop quantum gravity, emphasizing the ideas and results needed for the present paper; this hopefully makes the paper accessible for a wide audience.  In section 3 we introduce the topic of {\it categorical holography} which provides a natural setting for bullk/boundary relationships in loop quantum gravity, while in section 4 we review the recently understood relationship between the dynamics of quantum gravity and thermodynamics.  This allows us to express an analogue of the Ryu-Takkayanagi relation in section 5, and derive it in section 6.  We follow that with a statement of a conjecture of another form of the relation in section 7, after which the concluding section closes the paper.

\section{The basic assumptions of loop quantum gravity}

Loop quantum gravity is based on a few simple physical ideas.

\begin{enumerate}

\item{} The dynamical coordinate of general relativity is
a connection: the Ashtekar connection\cite{Abhay,books}.  
This means that general relativity is a diffeomorphism invariant gauge theory.

\item{}The degrees of freedom of a quantum
gauge theory are quanta of electric flux, which can be represented as holonomies and interpreted as loops or  strings.
This is also called the dual superconductor  picture.  It leads to the loop representation\cite{looprep,area}.

$1$ and $2$ together impy together that area and volume have discrete spectra\cite{area}.

\item{}General relativity is a
constrained topological field theory\cite{BC,FK,EPRL}.   This means that general relativity arises from a topological field theory by the
imposition of constraints. These are called the {\it simplicity constraints}\cite{BC,FK,EPRL}.  They are as simple as possible, in that they are quadratic, non-derivative local equations.
These break two degrees of freedom per point of the gauge invariance, leading to the emergence of the two physical degrees of freedom per 
point of general relativity.

The particular $TQFT$  general relativity is derived from is $BF$ theory.   
The action on a four manifold ${\cal M} = \Sigma \times R$ is\cite{CDJ}
\f
S= \int_{\cal M} B^i \wedge F_i - \frac{\Lambda}{2} B^i \wedge B_i -\phi_{ij} B_i \wedge B_j
- \frac{k}{4\pi } \int_{\partial {\cal M}} Y_{CS} (A)
\ff
Here $F^i$ is the curvature of an $SU(2)$ connection, $A^i$ and $B^i$ is a triad of two forms\footnote{How the equations of general relativity follow from this action is reviewed in \cite{positive}.}.
The boundary action is the Chern-Simons invariant of the pull back of the connection to the boundary
$\partial {\cal M} = S^2 \times R$.   
Note that without the term in $\phi_{ij}$, which is a traceless symmetric tensor in $ij$, this would be topological $BF$ theory.  The $\phi_{ij}$ are Lagrange multipliers whose equations of motion lead to the simplicity constraints.
\f
B^i \wedge B^i - \frac{1}{3} \delta^{ij} B^k \wedge B_k =0
\ff
The solutions of this is that $B^i$ is the self-dual two form of some metric; and that metric satisfies the Einstein equations, 
by virtue of the $A^i$ and $B^i$ field equations. 

 For our purposes we focus on the boundary conditions required for the action to be functionally differentiable, as these are central for the bulk-boundary relation.  The variational principle imposes boundary conditions which are given by\cite{linking}
\f
(F^i - \frac{\Lambda}{2 \pi} B^i )|_{\partial {\cal M}}=0
\label{sd}
\ff
These are called the self-dual boundary conditions because they impose that the pull back of the field strength into the boundary is proportional to the self-dual two form, also pulled back.  Note that were that relation imposed on all six components of the field strength it would imply that the spacetime was deSitter or anti-deSitter, depending on the sign of $\Lambda$. So this  can be seen as a topological form of an asymptotic deSitter or anti-deSitter boundary condition.

\item{} The cosmological constant is implemented by means quantum  deformation of  the representation theory used for the spin network labels. In the Euclidean case this involves $q$ at a root of unity, where  the level  is given by\cite{linking,qdef}
\f
k=\frac{6 \pi}{G\hbar \Lambda}
\ff
For the Lorentzian case we describe here, $q$ must be real so we have instead\cite{qrep,qsf}
\f
-\imath k=\ln (q) = \frac{6 \pi}{G\hbar \Lambda}
\ff

This implies that quantum general relativity is quantum $BF$ theory with defects\cite{linking,defects}.  These are punctures on the two dimensional spatial boundary, $\cal B$ and spin networks in the three dimensional bulk, $\Sigma$.  The self-dual boundary conditions 
force the free ends of the spin networks to be attached to punctures in the boundary, with the same labels.

\item{}Topological field theories in adjacent dimensions are related by a ladder of dimensions\cite{Louis1},
\f
BF_{3+1} \rightarrow CS_{2+1} \rightarrow WZW_{1+1}
\ff

The different dimensions are tied together by the self-dual boundary conditions(\ref{sd})

\item{} The simplicity constraints that turn $BF$ theory into general relativity have a simple geometrical interpretation.
The simplicity constraint can be expressed as a constraint on the representation theory used to label spin networks and spin foams. We first review this structure for the classical case and then describe its extension to quantum spin networks, labeled by representations of quantum groups.   The representations employed in the bulk are a subset of the representations of  $SL(2,C)$, which satisfy the following
{\it simplicity constraint},
\f
<{\cal S}>  =  < ( \hat{K}_z - \gamma \hat{L}_z )>=0
\label{S}
\ff
where $\gamma$ is the Immirzi parameter\footnote{An analogue of the $\theta$ parameter in $QCD$, that also measures the area gap.}, and $ \hat{K}_z$ and $ \hat{L}_z$ are the generators in the representation of boosts and rotations.  These require that the $SL(2,C)$ representations are images of $SU(2)$ representations under the map (\ref{Y})
\f
Y_\gamma : {\cal V}_j  \rightarrow {\cal V}_{j,r=\gamma (j+1)}.
\label{Y}
\ff
The simplicity constraints also have a compelling physical interpretation in terms of thermodynamics, which complements their elegant mathematical interpretation, and is the subject of section 3, below.



\end{enumerate}

\begin{figure}[H]
\begin{center}
\includegraphics[width=.8 \textwidth]{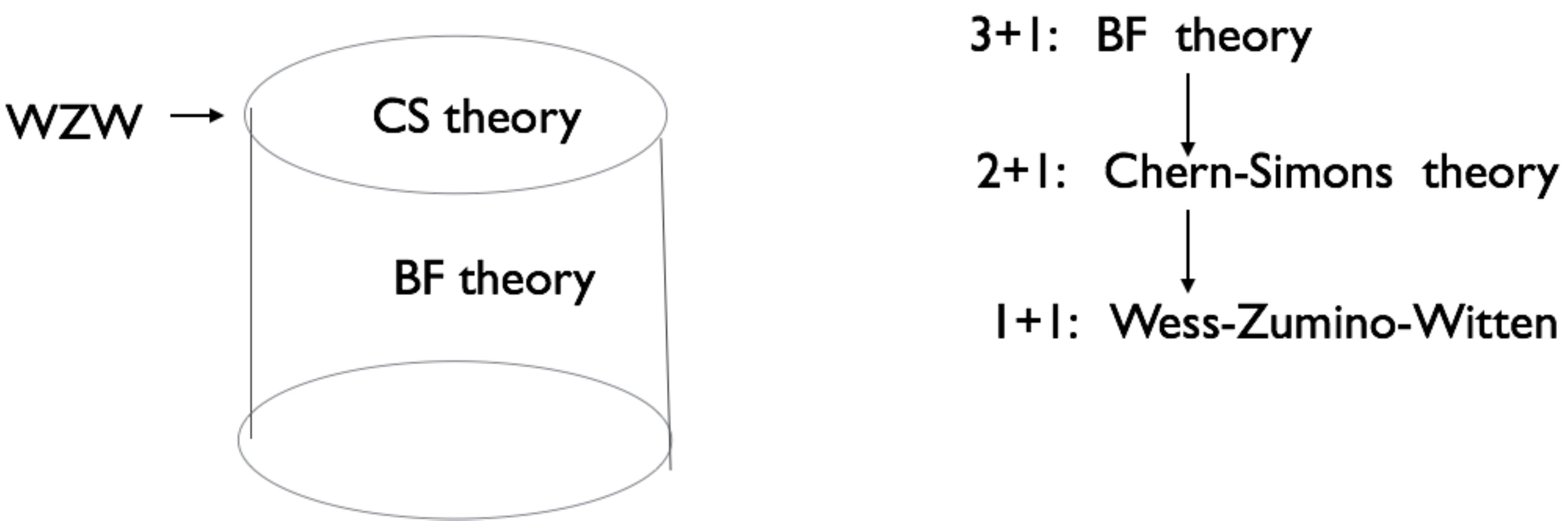}
\end{center}
\caption{Ladder of dimensions}
\label{fig1} 
\end{figure}

\begin{figure}[H]
\begin{center}
\includegraphics[width=.8 \textwidth]{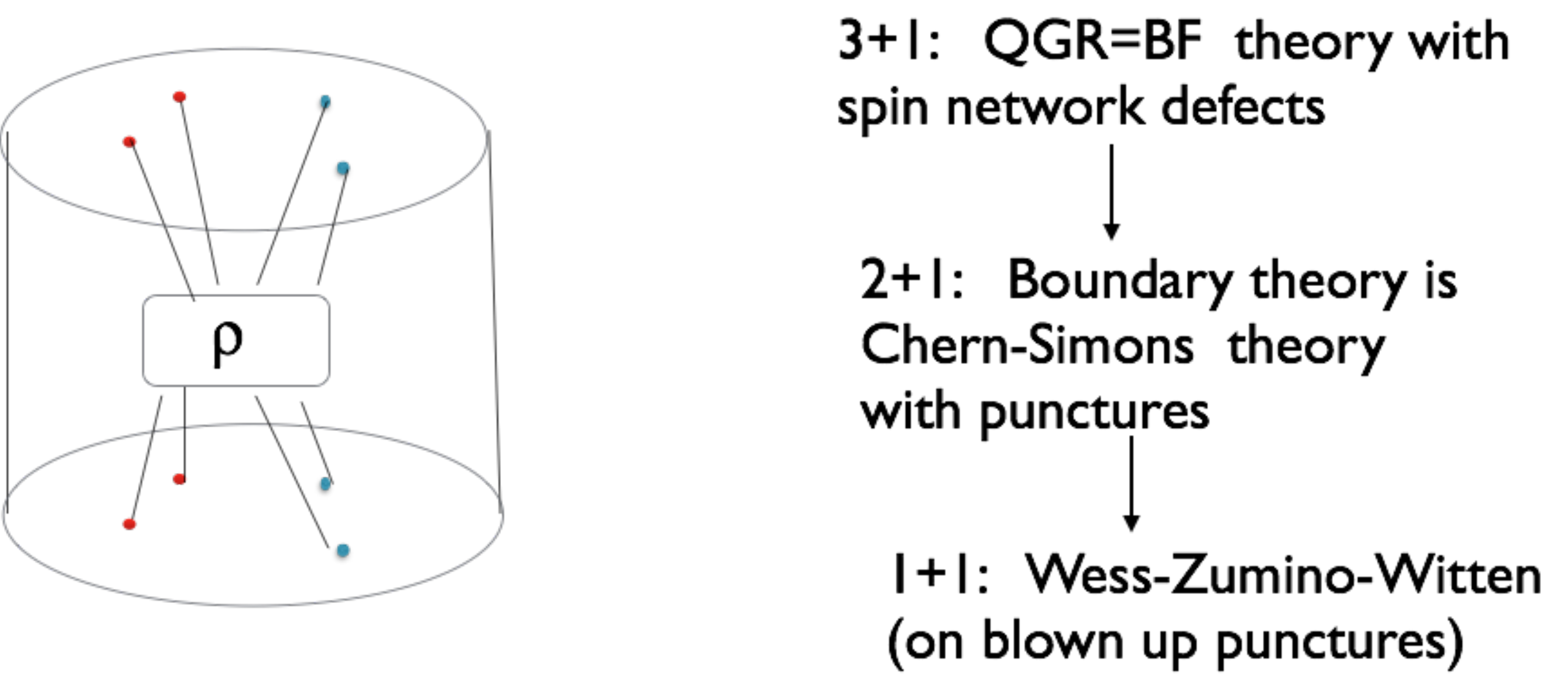}
\end{center}
\caption{The quantum ladder of dimensions reveals quantum general relativity as a $TQFT$ with defects.}
\label{fig2} 
\end{figure}

\subsection{Extension to quantum group representations}

So far we followed \cite{Eugenio1} in defining the simplicity map $Y_\gamma$, (\ref{Y}) for the ordinary groups, $SU(2)$
and $S0(3,1)$.   But the connection to Chern-Simons theory as well as the form of the bulk-boundary map requires that the representations be quantum deformed.  Hence we take the boundary states to be labeled by the representation theory
of $SU_q (2)$ and the bulk spin networks to be labeled by the infinite dimensional unitary representations of 
$SL_q (2,C)$.  As in the classical case, the former are labeled by an half integer, $j$ while the latter
are labeled by a pair $(j,r)$ where $r$ is a real number \cite{qrep,qsf}.  The latter exist in the case of $q$ real, which means we must take for the level 
\f
\ln (q) = \frac{6\pi}{G \hbar \Lambda}
\label{lnq}
\ff
the simplicity map is now,
\f
Y_\gamma^q : {\cal V}_j^q  \rightarrow {\cal V}_{j,r=\gamma (j+1)}^q.
\label{Yq}
\ff

\section{Categorical holography.}

These mathematical structures reveal a beautiful bulk-boundary correspondence, which provides a non-perturbative realization of holography.  This might be called {\it categorical holography} and was first described by Louis Crane\cite{Louis1}.

\begin{itemize}

\item{}To every punctured 2-surface, $B$, with labels, $j_i$ we associate a Hilbert space, ${\cal H}_{B  j_i} $ as shown 
in Figure \ref{fig3}.

\begin{figure}[H]
\begin{center}
\includegraphics[width=.8 \textwidth]{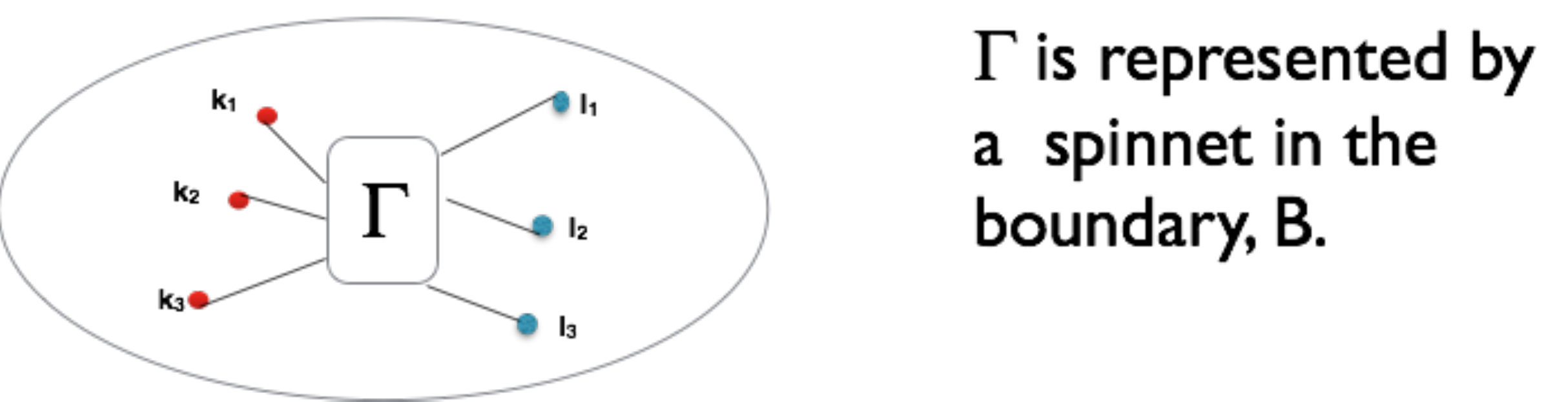}
\end{center}
\caption{To every punctured 2-surface, $B$, with labels, $j_i$ we associate a Hilbert space, ${\cal H}_{{\cal B}  j_i} $}
\label{fig3} 
\end{figure}
This is the Hilbert space of Chern-Simons theory at level $k$ on the punctured surface $\cal B$, with punctured labeled in the 
representations of $SU_q (2)$.  It is spanned by quantum $SU_q(2)$ spin networks on $\cal B$ with ends attached to the punctures.

\item{}To the surface with reversed orientation we associate the dual  Hilbert space, ${\cal H}_{{\cal B}  j_i}^\dagger $.

\item{}Consider a three dimensional manifold $\Sigma$, whose boundary is $\cal B$.  On it we define for each set
of labeled punctures on ${\cal B}$ a bulk Hilbert space, ${\cal H}_{\Sigma, j_i }$, spanned by a basis of quantum 
$SL_q (2,C)$ spin networks with ends on the punctures.  We identify the puncture labels from $SU_q (2)$ with the 
puncture labels for simple representations of $SL_q (2,C)$ with the map $Y_\gamma^q$, (\ref{Yq}).

\item{} A bulk boundary correspondence is gotten by taking a quantum spin network in the boundary and suspending it into the bulk from the punctures.  This suspension is well defined because quantum spin networks in the two dimensional boundary distinguish over-crossings from under-crossings.  We then use the map $Y_\gamma^q$ to transform the $SU_q(2)$ spin network labels to those of simple 
$SL_q(2,C)$ representations.  This  gives a holographic map
\f
{\cal T}_\gamma^q : {\cal H}_{{\cal B}  j_i} \rightarrow {\cal H}_{\Sigma, j_i }
\ff

\begin{figure}[H]
\begin{center}
\includegraphics[width=.6 \textwidth]{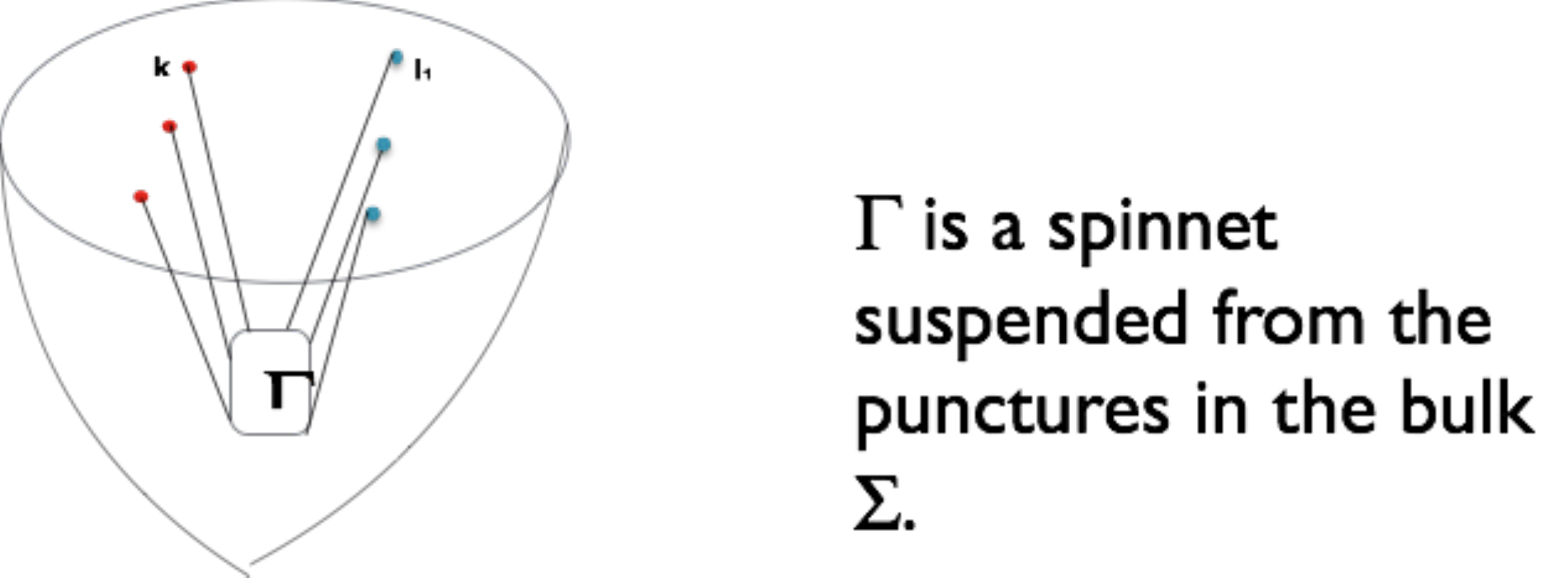}
\end{center}
\label{fig4} 
\end{figure}

\begin{figure}[H]
\begin{center}
\includegraphics[width=.6 \textwidth]{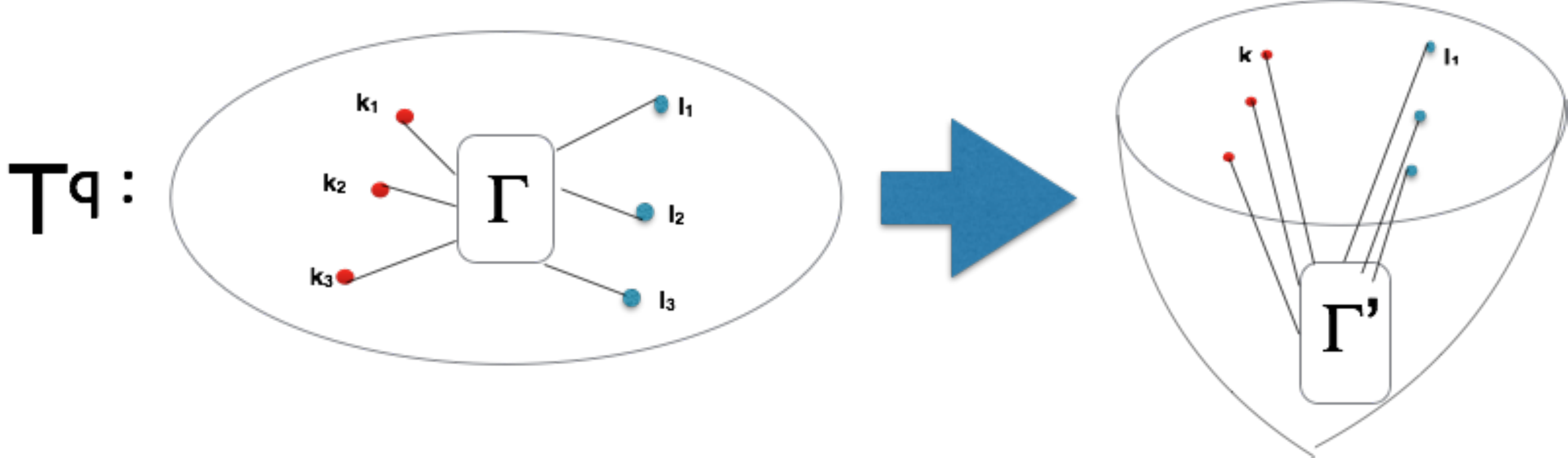}
\end{center}
\caption{The holographic map from the boundary into the bulk Hilbert space.}
\label{fig4b} 
\end{figure}

\item{}To each operator on the boundary Hilbert space we associate a cobordism, ie a cylinder of one higher dimension.

\begin{figure}[H]
\begin{center}
\includegraphics[width=.5 \textwidth]{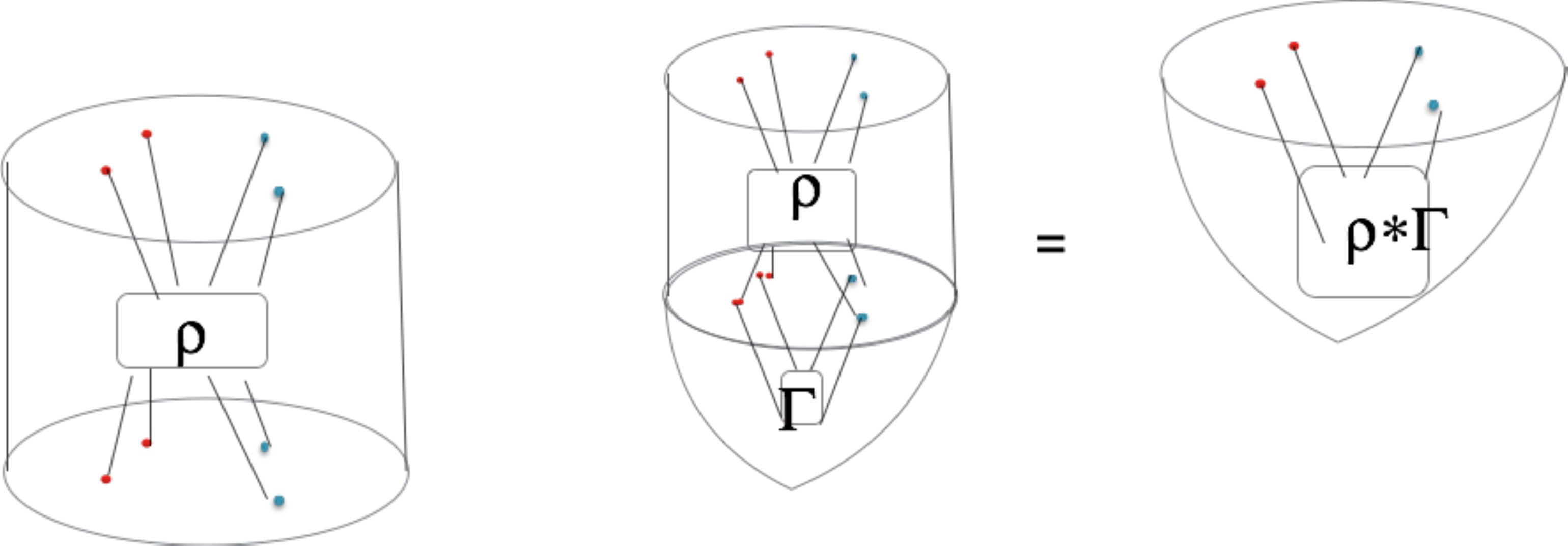}
\end{center}
\label{fig4c} 
\end{figure}

\item{} Trace is represented by closing the cylinder.
\begin{figure}[H]
\begin{center}
\includegraphics[width=.3 \textwidth]{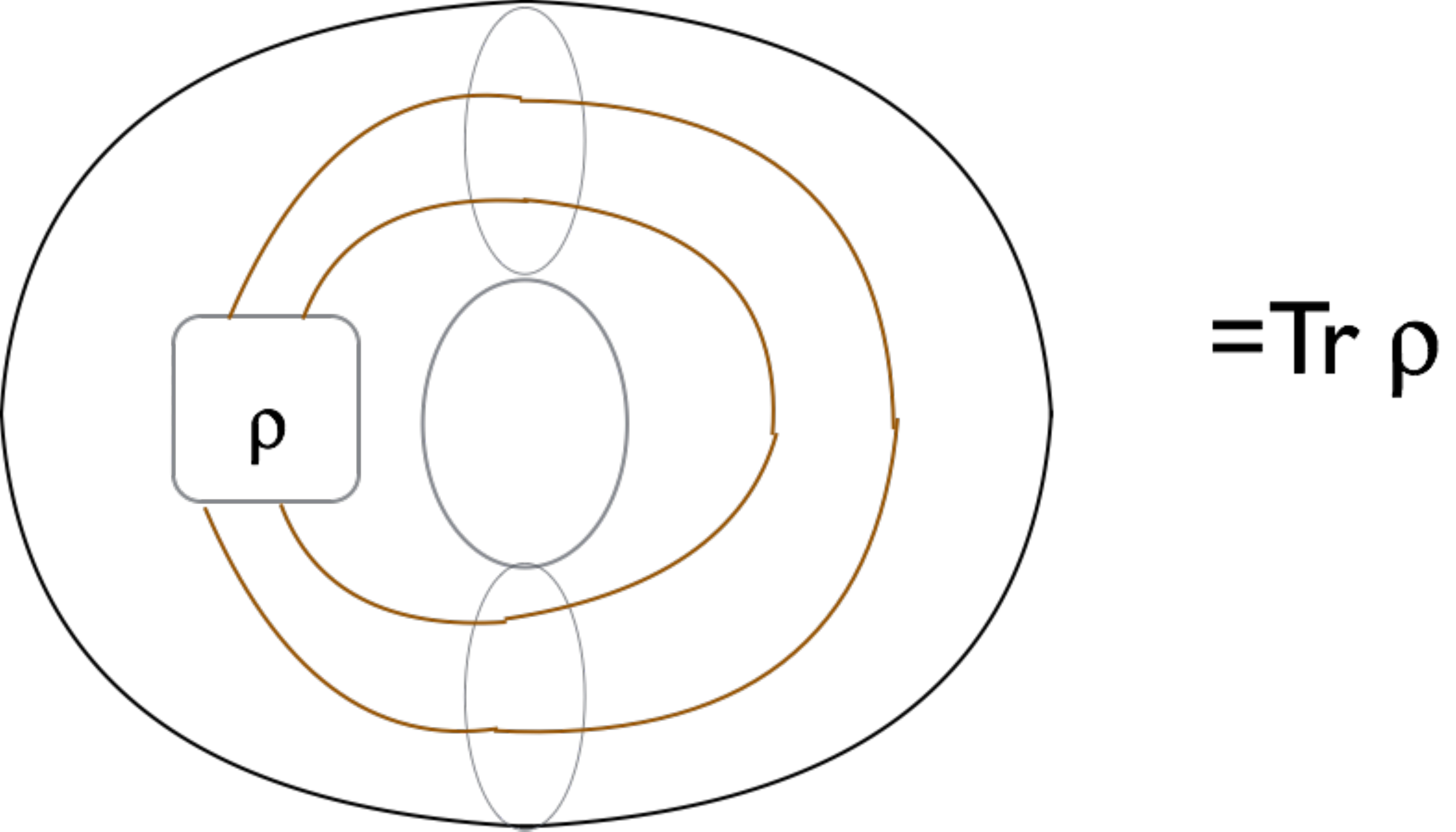}
\end{center}
\label{fig5} 
\end{figure}
\end{itemize}


\section{The simplicity constraint expresses the first law of thermodynamics}

We will next need the thermodynamic interpretation of the simplicity constraints, (\ref{S}).  This is presented in \cite{carloetal,FGP,Eugenio1,thermost}.

Note that the boost Hamiltonian acts within the $SL(2,C)$ representations as,
\f
\hat{H}_b = \hbar \hat{K}_z 
\ff

while the boost temperature has dimensions of $\hbar$ because the boost is dimensionless, and has the 
universal value
\f
\beta = \frac{2 \pi}{\hbar} .
\ff
Recall also that the area operator is related to a rotation generator by
\f
\hat{A}= 8 \pi \gamma G \hbar |\hat{L}_z |
\ff
Combining these and 
taking the expectation value yields the first law of thermodynamics
\f
\beta <\hat{H}_b >= S_T = \frac{<\hat{A}>}{4G\hbar}, 
\ff
Note that the $\hbar$'s and $\gamma$'s cancel, so we get a classical expression for the expectation values, which
expresses what has been called the first law of spacetime dynamics\cite{CT,MP,JP,BW,ls-bh}.
\f
H_b = \frac{A}{8 \pi G}
\ff
This is also called the Carlip-Teitelbiom relationc\cite{CT,MP,JP,BW,ls-bh}.  We may note that from it the positivity of $H_b$  follows from that of the area; this is possibly related to \cite{ooguri}.

\section{Analogues of the  Ryu-Takayanagi relation}

We next proceed by dividing the punctured boundary $\cal B$ into two regions, $A$ and its complement $B= \neg A$, 
separated by a circle $\gamma$ in $\cal B$.  $A$ contains $n_A$ punctures with labels $j_I$ while
$B$ contains $n_B$ punctures with labels $k_I"$.  We assume no punctures lie on $\gamma$.

Consider a quantum spin network in $\cal B$.  This will have $p$ edges
crossing $\gamma$ with representations $l_1 , \ldots , l_p$.  There is a Hilbert space ${\cal H}_{A, \{ l \}}$
in region $A$ spanned by spin networks that end at the $n_A$ punctures and at the $p$ edges that cross
$\gamma$, and similarly for $B$.  We then may decompose the original Hilbert space as
\f
{\cal H}_{{\cal B} \{ j \}  \{ k \} } = \sum_{p=0}^\infty \sum_{\{ l \} }  {\cal H}_{ A, \{ j \}   \{ l \} }   \otimes {\cal H}_{B, \{k \}  \{ l \} }
\ff
\begin{figure}[H]
\begin{center}
\includegraphics[width=.3 \textwidth]{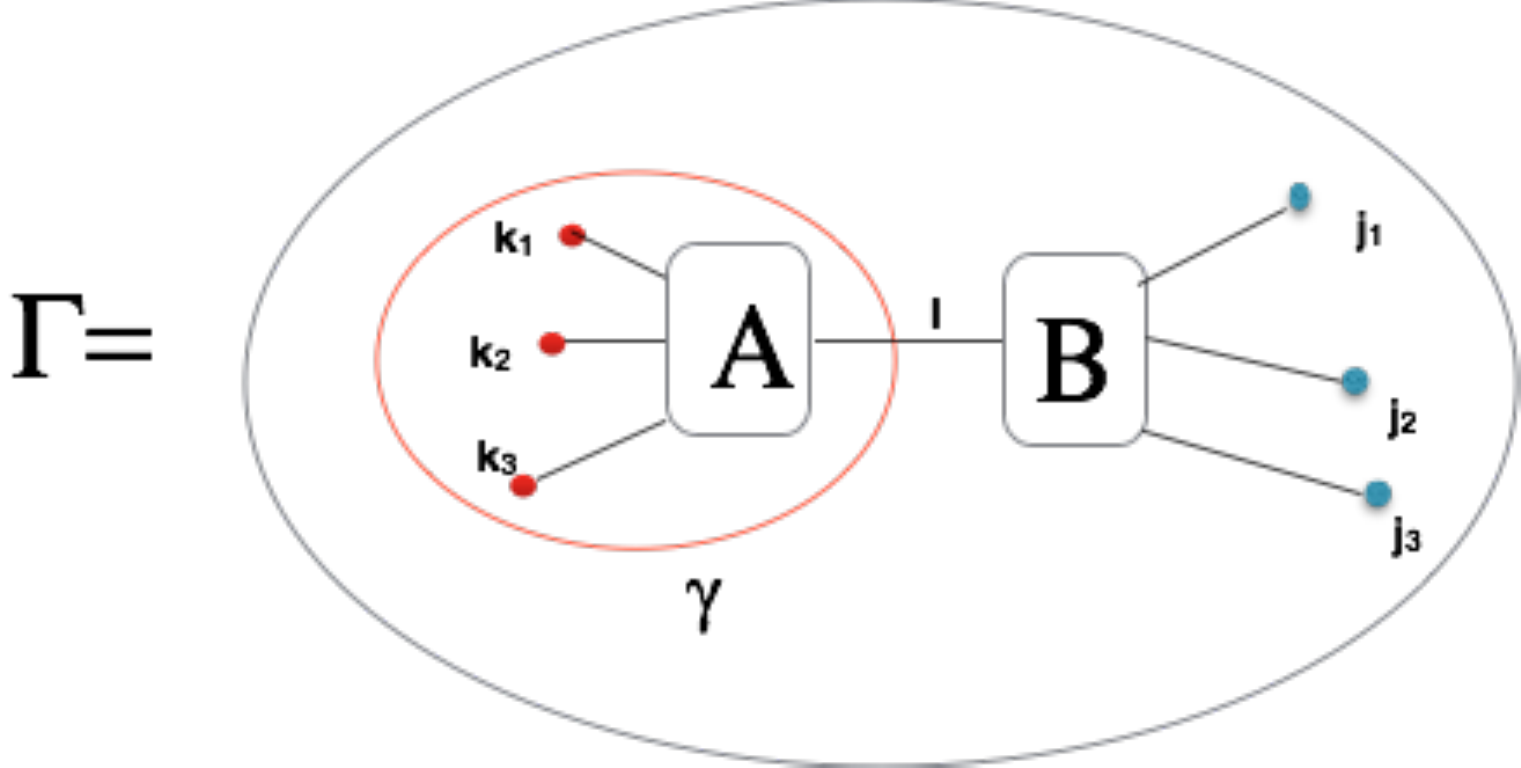}
\end{center}
\caption{Dividing the boundary theory into two subsysetems.}
\label{fig7} 
\end{figure}

\begin{figure}[H]
\begin{center}
\includegraphics[width=.3 \textwidth]{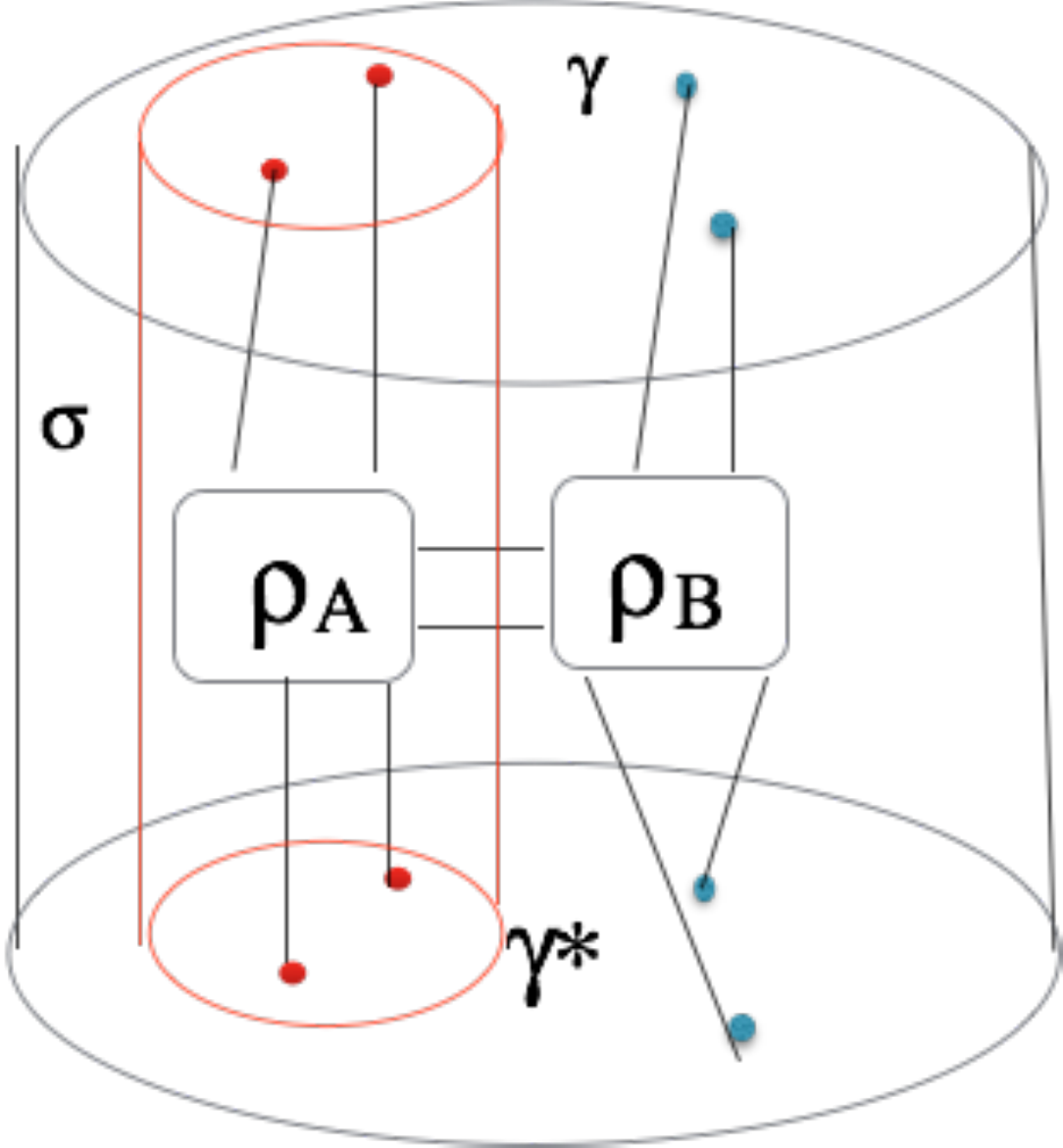}
\end{center}
\caption{Dividing the density matrix of the boundary theory into two subsysetems.}
\label{fig8} 
\end{figure}

We will also need corresponding the decomposition of density matrices.  Note that this takes place in
the cylinder ${\cal B} \times I$ as indicated in figure \ref{fig8}.

We now define an thermal density matrix in ${\cal H}_{{\cal B} \{ j \}  \{ k \} } $.  This is defined by an operator 
in ${\cal H}_{{\cal B} \{ j \}  \{ k \} } $ which can be represented as a cobordism embedded in the cylinder
$A \times R$, as shown in Figure \ref{fig8}.  The cylinder ends are a copy of the punctured surface $A$ and its dual
$\tilde{A}$

\begin{figure}[H]
\begin{center}
\includegraphics[width=.3 \textwidth]{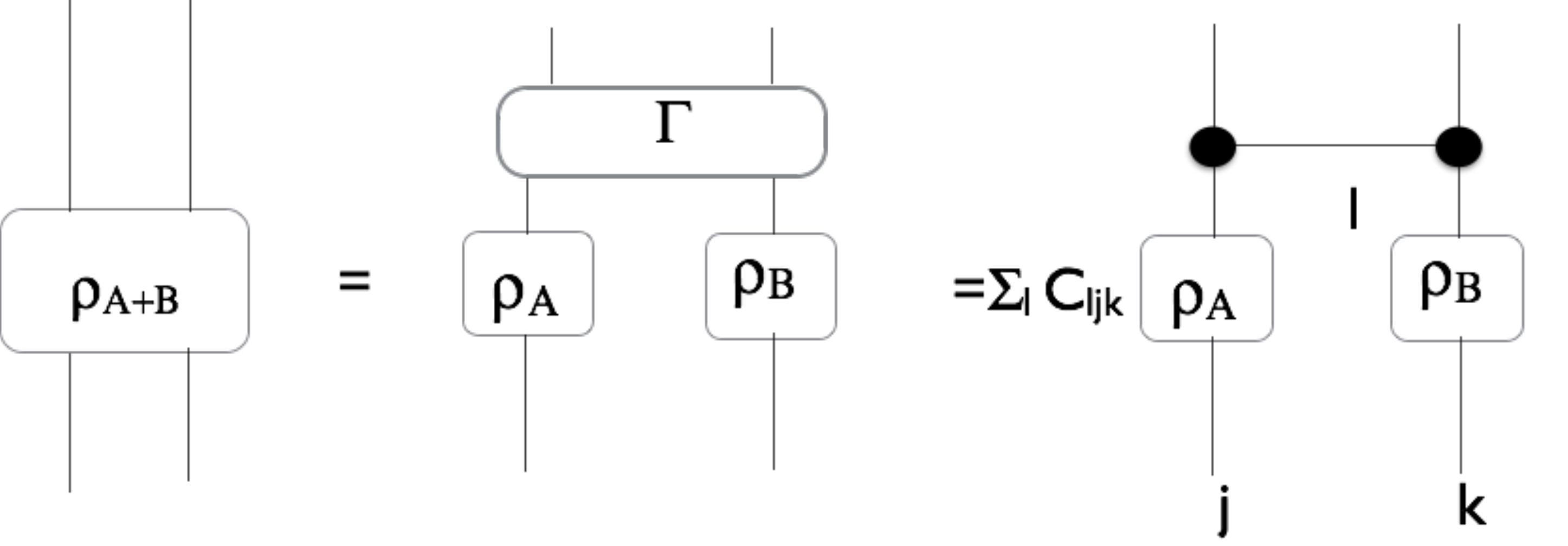}
\end{center}
\caption{An example of an entangled density matrix.}
\label{fig9} 
\end{figure}
\begin{figure}[H]
\begin{center}
\includegraphics[width=.3 \textwidth]{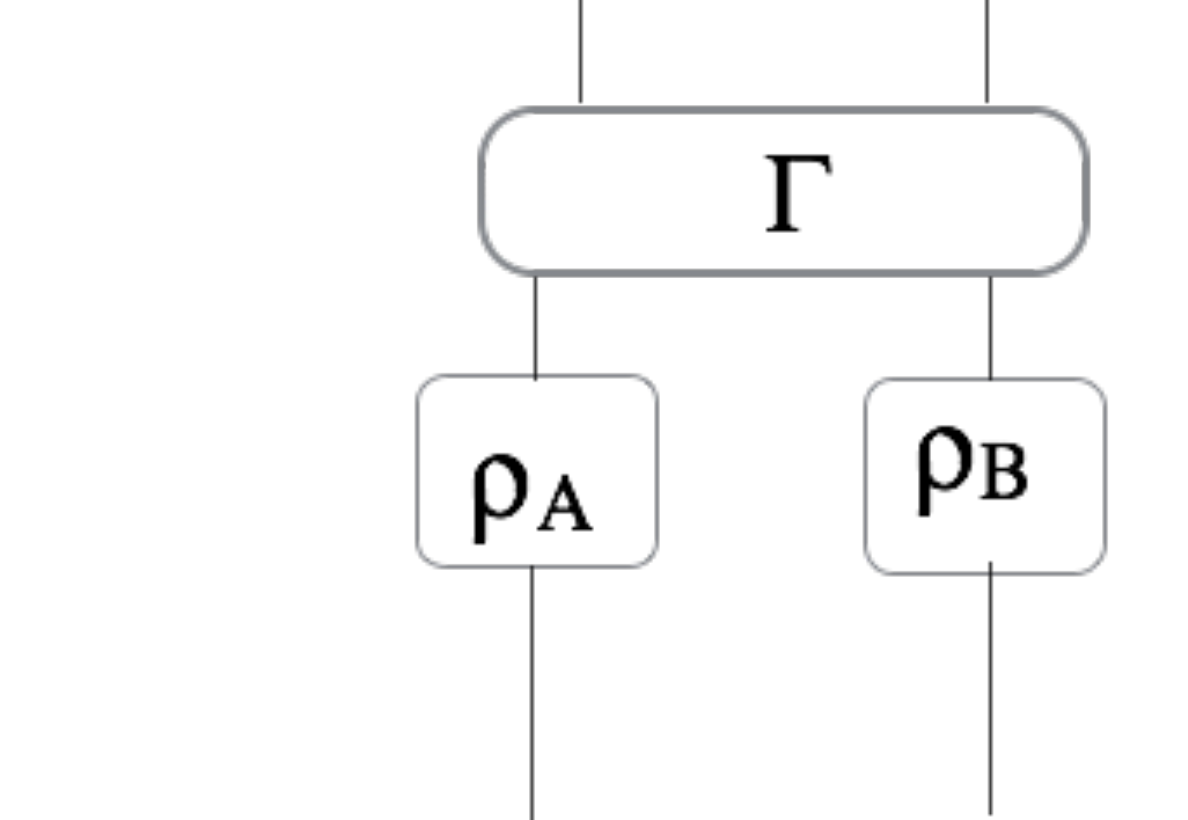}
\end{center}
\caption{A general set of entangled thermal states.}
\label{fig10} 
\end{figure}

This has the form of a product of spin network edges connecting punctures in $A$ with the dual punctures in
$\tilde{A}$.  Inserted into each edge with spin $j$ is a density operator\cite{Eugenio1}
\f
\rho_j = \frac{1}{Z} e^{-\beta \hat{H}_{j}}
\ff
This is illustrated in Figure \ref{fig10}.

\begin{figure}[H]
\begin{center}
\includegraphics[width=.3 \textwidth]{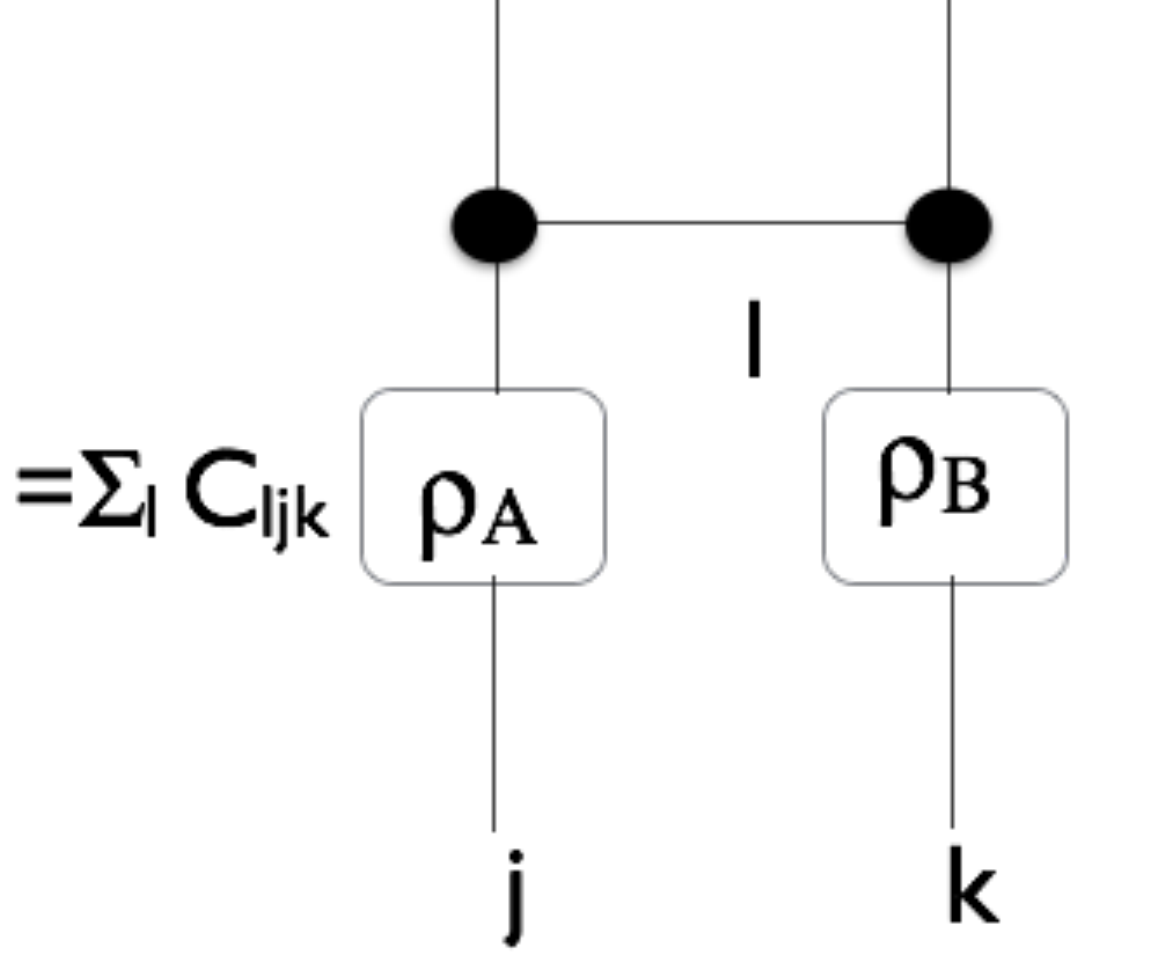}
\end{center}
\caption{A more specific set of entangled thermal states.}
\label{fig10b} 
\end{figure}

We now consider a class of entangled-thermal states, shown in Figure \ref{fig10}.
A specific example is shown in Figure \ref{fig10b}.

We next define the thermal entropy of this entangled state as 
\f
S_T = \beta Tr( \rho_{A+B} H_b )  = \beta  < H_b    >
\ff

Let us consider now surfaces, $\sigma$, in the cylinder, spanning the loops $\gamma$ and its dual $\tilde{\gamma}$
\f
\partial \sigma = \gamma -  \tilde{\gamma}
\ff
This will intersect various edges in the bulk.  For each such surface there is an average area defined by
\f
< A[\sigma ]> = Tr \rho_{A+B} \hat{A}[\sigma ] 
\ff
We will show below that there is, up to topology, a minimal area surface $\sigma_{min}$ spanning a loop 
$\gamma \in {\cal B}$ and its image $\tilde{\gamma}$.   One can define an operator
\f
\hat{A}_{min}(\gamma )= \hat{A}[\sigma_{min}]_{\partial \sigma_{min} = \gamma -\tilde{\gamma}}
\ff
by defining its action in the spin network basis.
  
Given this we will also show that
\f
S_T = \frac{<\hat{A}[\sigma_{min}]>}{4 G \hbar  }
\ff

This is a form of the the Ryu-Takayanagi relation.

\begin{figure}[H]
\begin{center}
\includegraphics[width=.7 \textwidth]{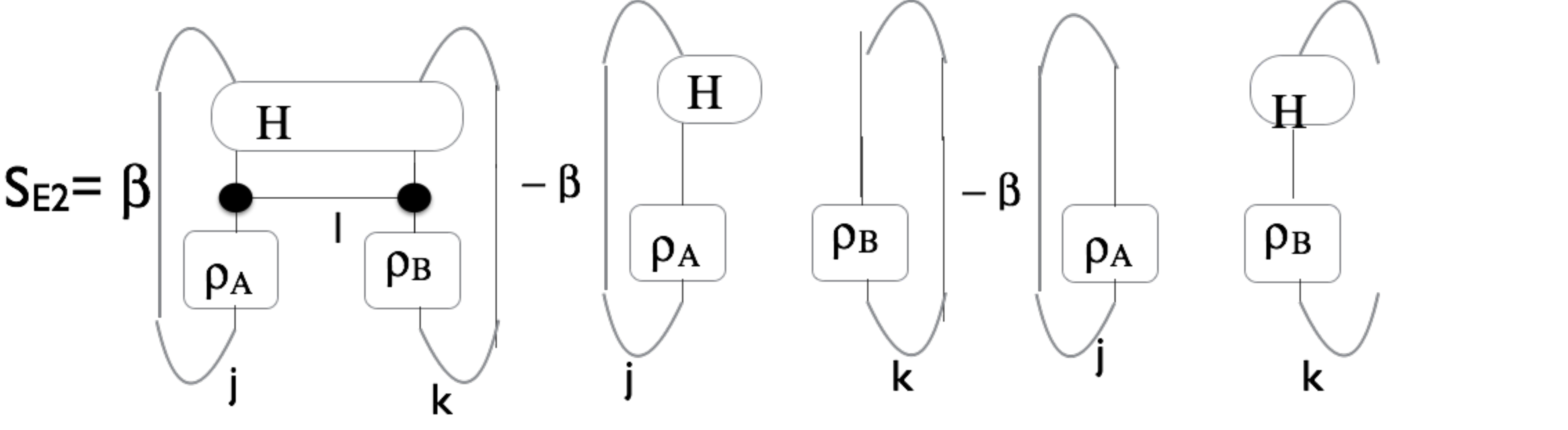}
\end{center}
\caption{One form of the entanglement entropy.}
\label{fig18e} 
\end{figure}

One might object that $S_T$ is not an entanglement entropy, it is a thermal entropy.  To get an entanglement entropy
one sould subtract of the entropy of the subsystems.  One way to do this
 for states of the form of Fig. (\ref{fig10}) is to take the difference between $S_T$ and  the entropy of the simple thermal product state of the two subsystems
\f
S_{E2}= S_T (\rho_{A+B}) - S_T(\rho_A) - S_T (\rho_B)
\ff
This is shown in Figure \ref{fig18e}.   
It is clear that $S_T(\rho_A)$ and $ S_T (\rho_B)$ do not depend on the area of any spanning surface.  Hence we have
\f
S_{E2} = \frac{<\hat{A}_{min}(\gamma )>}{4G\hbar}  + C
\ff
where $C$ is a constant that doesn't depend on the area of the surface $\sigma$.

This is an analogue of the Ryu-Takayanagi relation.  \section{The derivation}

\begin{figure}[H]
\begin{center}
\includegraphics[width=.5 \textwidth]{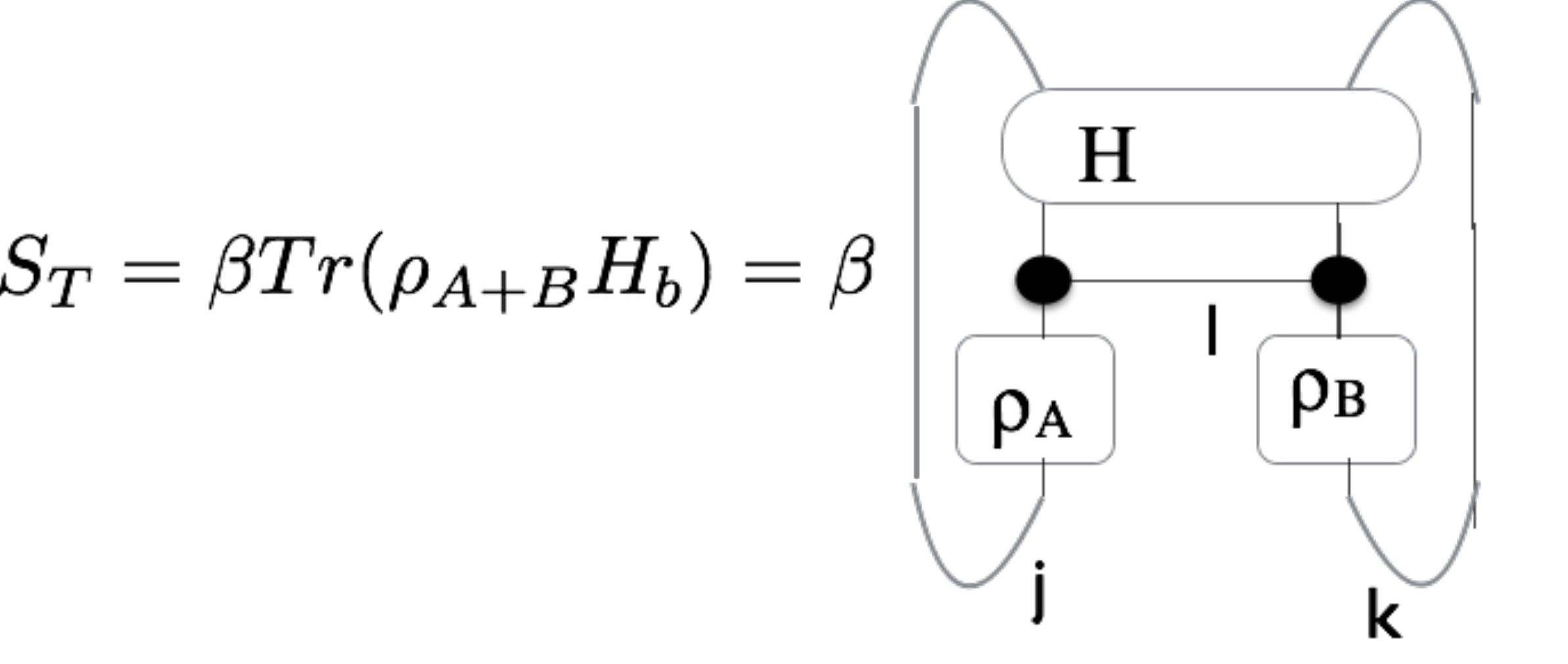}
\end{center}
\caption{The entropy $S_T$.}
\label{fig11st} 
\end{figure}

We want to compute
\f
S_T = \beta Tr( \rho_{A+B} H_b ) 
\ff
This is represented diagrammatically in Figure \ref{fig11st}.
For the specific form \ref{fig10b}, that gives
\begin{figure}[H]
\begin{center}
\includegraphics[width=.5 \textwidth]{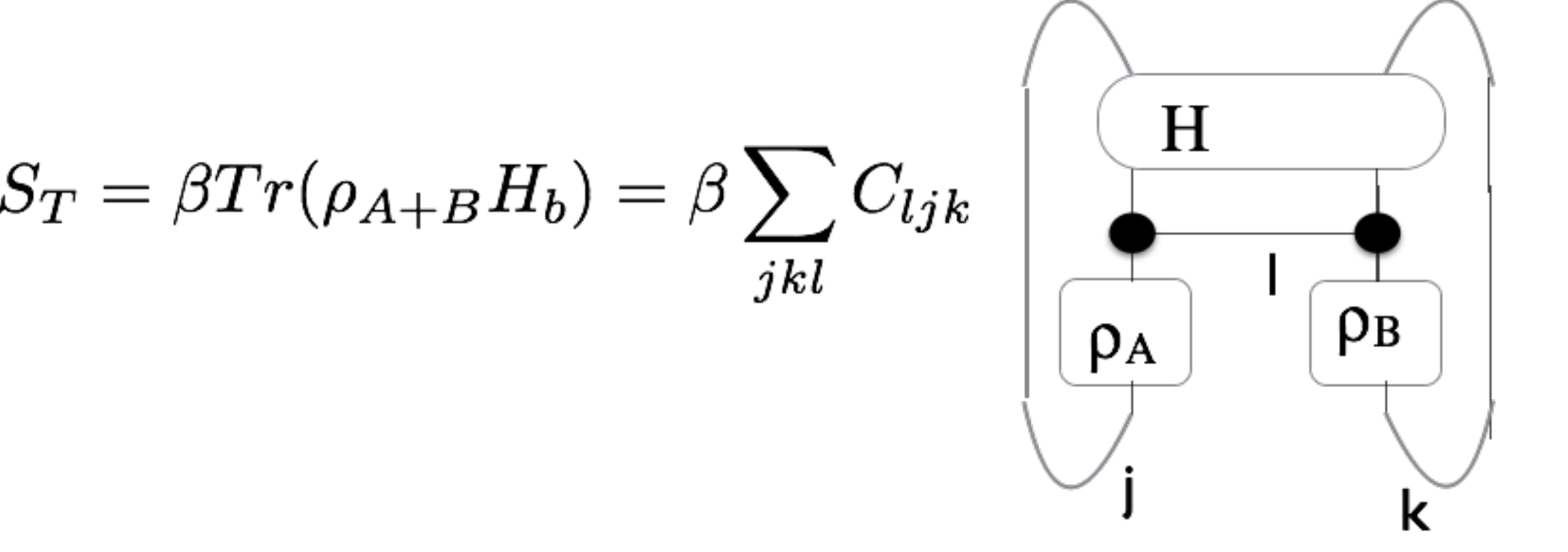}
\end{center}
\label{fig12} 
\end{figure}
Next we use the decomposition of the boost generator
\f
H_b = H_A \otimes I + I \otimes H_B
\ff
To find
\begin{figure}[H]
\begin{center}
\includegraphics[width=.5 \textwidth]{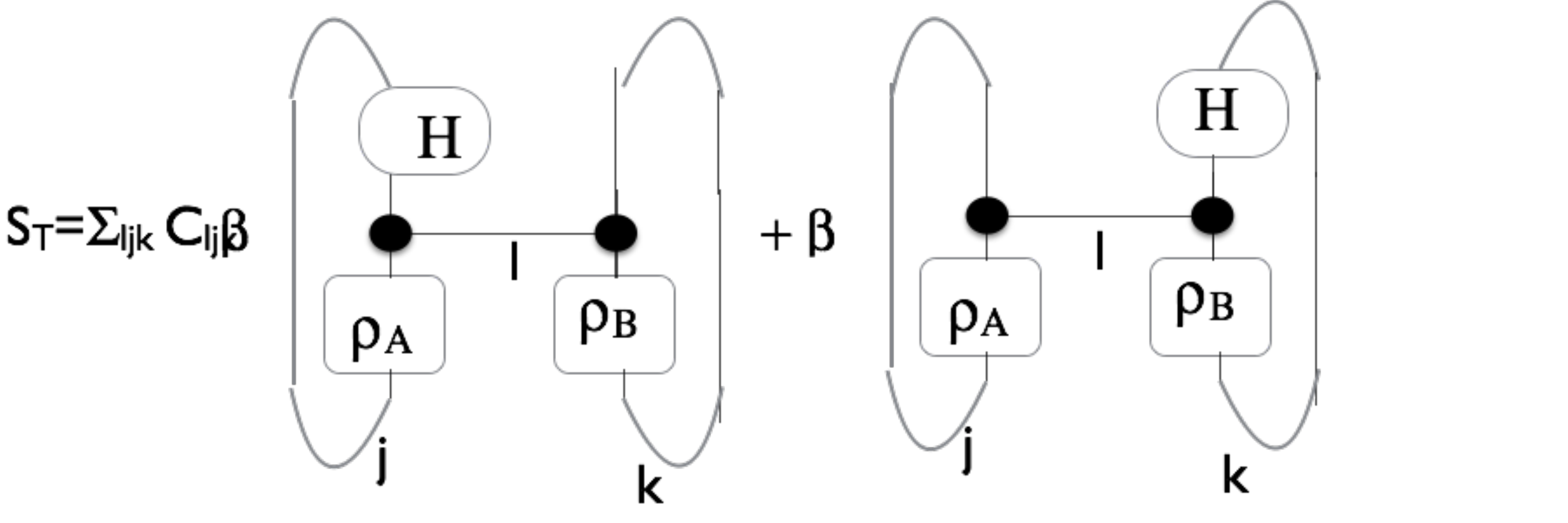}
\end{center}
\label{fig14} 
\end{figure}
We  next use the fact that $H_b$ generates a symmetry of the intertwiners,  
\begin{figure}[H]
\begin{center}
\includegraphics[width=.2 \textwidth]{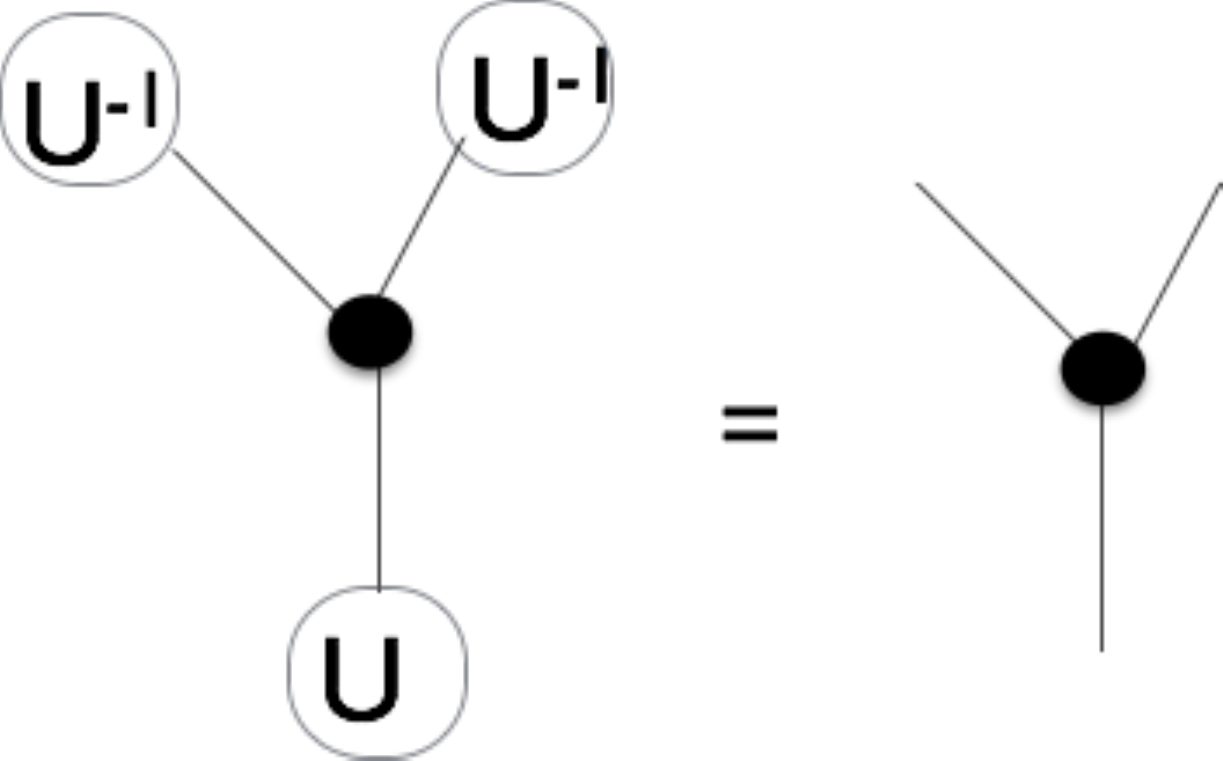}
\end{center}
\label{fig14b} 
\end{figure}
which implies
\begin{figure}[H]
\begin{center}
\includegraphics[width=.5 \textwidth]{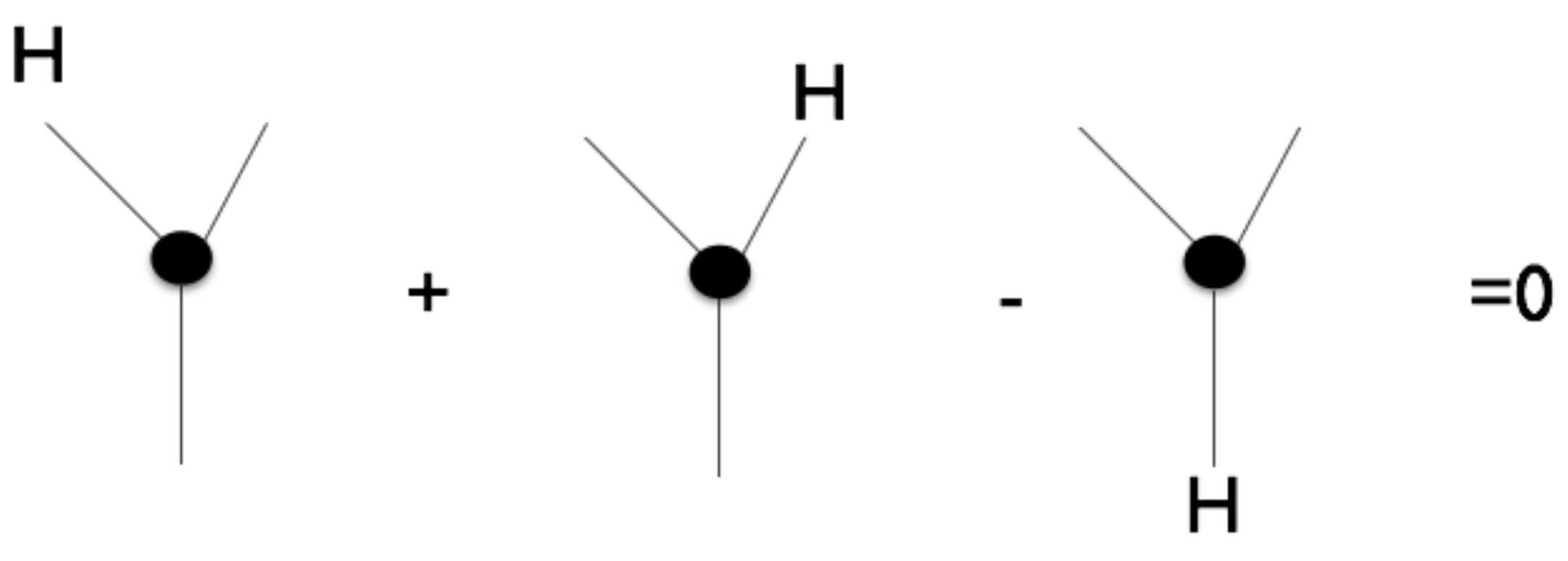}
\end{center}
\label{fig15} 
\end{figure}
This becomes
\begin{figure}[H]
\begin{center}
\includegraphics[width=.5 \textwidth]{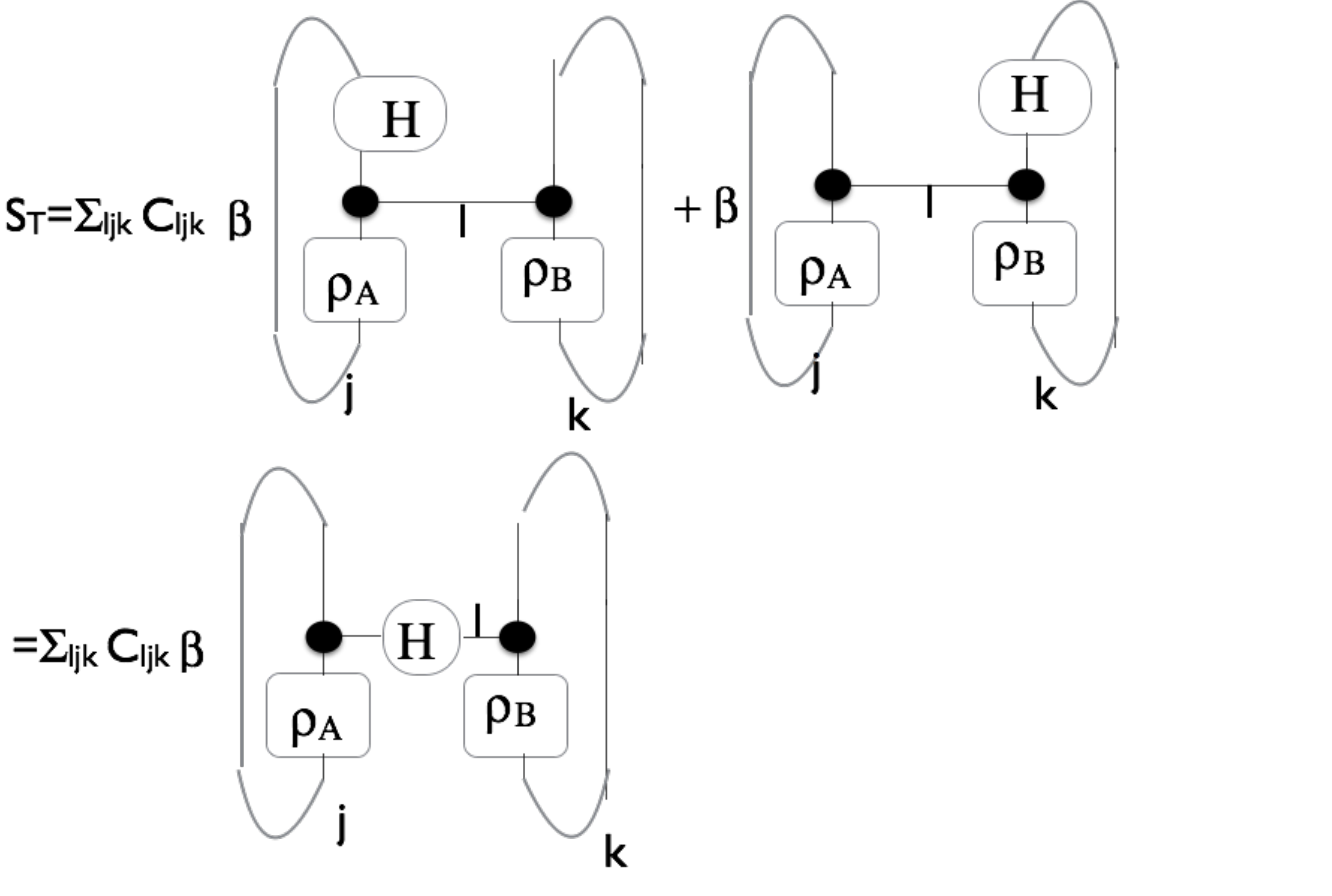}
\end{center}
\label{fig16} 
\end{figure}
We use the simplicity constraint (\ref{S}) to show that
\begin{figure}[H]
\begin{center}
\includegraphics[width=.6 \textwidth]{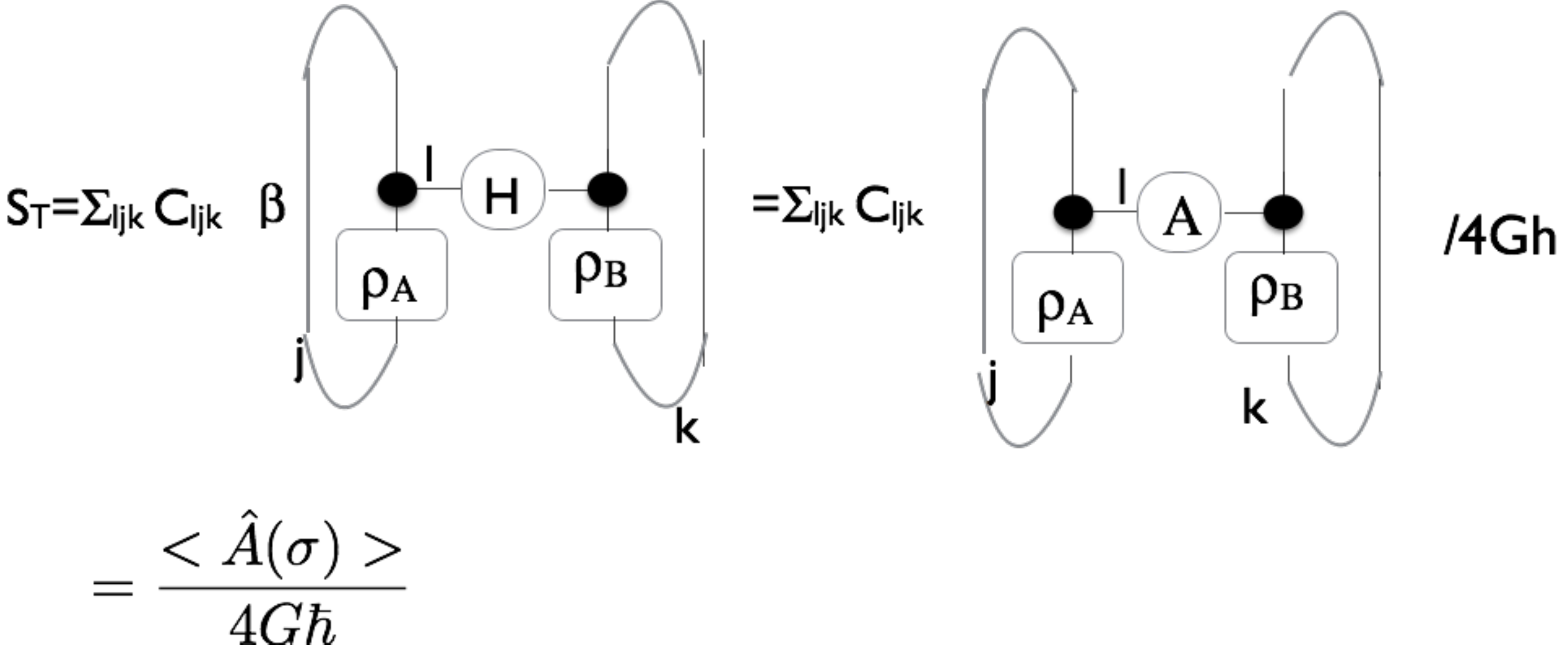}
\end{center}
\label{fig17} 
\end{figure}

\begin{figure}[H]
\begin{center}
\includegraphics[width=.4 \textwidth]{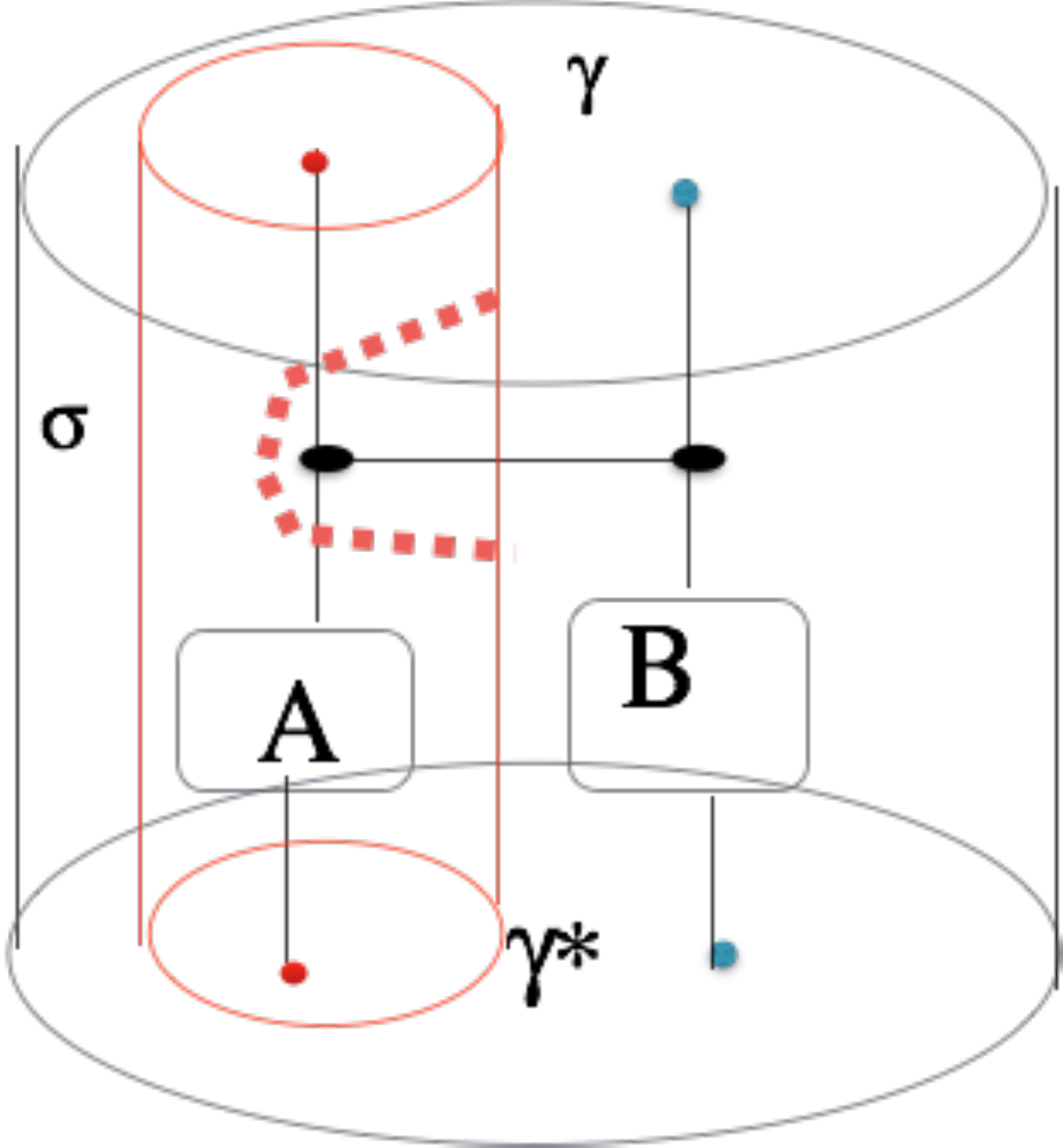}
\end{center}
\caption{Deformation of the surface $\sigma$.}
\label{fig18a} 
\end{figure}

There is one more step to the Ryu-Takayanagi relation which is to confirm that the cut we used to define the entropy comes from a minimal area surface.   We have to show that deforming the surface around a node as shown in Figure \ref{fig18a} increases the area.  
We have to show that
\f
<\hat{A}[\sigma_1 ]>  \leq<\hat{A}[\sigma_2 ]>
\ff
where $\sigma_1$ and $\sigma_2$ are defined by
\begin{figure}[H]
\begin{center}
\includegraphics[width=.3 \textwidth]{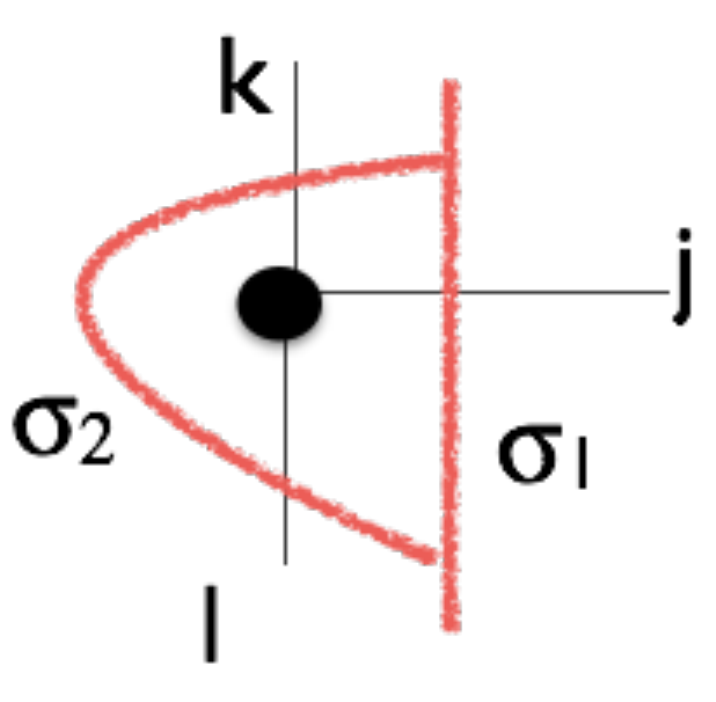}
\end{center}
\label{fig18} 
\end{figure}
This follows from the triangle inequality $j \leq k+l$, which expresses Gauss's law.  This implies that the area eigenvalues satisfy. 
\f
  A(j) = \leq A(k) + A(l)
\ff
So let us define the operator $\hat{A}_{min} [\sigma ]$ to return on eigenstates of $\hat{A}[\sigma ]$ the minimal area of a surface
spanning the same boundary $\partial \sigma = \gamma - \tilde{\gamma}$.  We then have shown
\f
S_T = \frac{<\hat{A}_{min}(\gamma )>}{4G\hbar}
\ff

\section{A conjecture}


We may instead define the entanglement entropy as the difference between the reduced and von Neumann entropies.
\f
S_E = S_{red} - S_{von N}
\ff
where
\f
S_{red} = - Tr \rho_A^{red} \ln \rho_A^{red}
\ff
and 
\f
S_{von N} = - Tr \rho_{A+B} \ln \rho_{A+B}
\ff
We conjecture that
\f
S_E = \frac{<\hat{A}_{min}(\gamma )>}{4G\hbar}  + C
\ff

\section{Conclusions}

We have found an analogue of the  Ryu-Takayanagi relation for loop quantum gravity, and conjectured another form of the result.
Let us recall the inputs to the derivation.

\begin{itemize}

\item{}The result is built on the basic pillars of loop quantum gravity: the recognition of gravity as a diffeomorphism invariant gauge theory, the duality of quantized gauge fields with extended objects, and in particular the identification of area with electric flux, the close relationship with topological $BF$  theory, and the resulting categorical holographic structure based on the ladder of dimensions, with a natural Chern-Simons boundary theory.

\item{}We impose a non-perturbative version of asymptotically AdS boundary conditions, expressed as the condition that the curvature pulled back to the boundary be self-dual

\item{}We represent the density matrix by spin-network in the bulk, following the framework of categorical holography.

\item{}We assume an entangled thermal-boosted state.

\item{}We use invariance of intertwiners to move symmetry generators around.

\item{}We use the simplicity constraint to relate the boost hamiltonian to area and hence to entropy, thus incorporating the first law of thermodynamimcs.

\end{itemize}

In the near future we hope to continue to deepen our understanding of the relationship between gravity, thermodynamics and the quantum, in the search for a deeper expression of the holographic principle.  One important step would be to understand the relationship between the emergence of  the Einstein equations from thermodynamics\cite{Ted95,Ted2015,thermost} and the kind of holographic identities derived here.  Related to this would be to find a quantum gravity setting to express the positivity of the boost hamiltonian to that of entropy\cite{ooguri}, given that the former is a consequence of the Carlip-Teitelbiom relation\cite{CT,MP,JP,BW,ls-bh}.

\section*{ACKNOWLEDGEMENTS}

It is a pleasure to thank Bianca Dittrich, Laurent Freidel,  Marc Geiler, Aldo Riello, Carlo Rovelli and Wolfgang Wieland for comments and discussion.  

This research was supported in part by Perimeter Institute for Theoretical Physics. Research at Perimeter Institute is supported by the Government of Canada through Industry Canada and by the Province of Ontario through the Ministry of Research and Innovation. This research was also partly supported by grants from NSERC, FQXi and the John Templeton Foundation.

\end{document}